# A Two-Stage Trip Inference Model of Purposes and Socio-Economic Attributes of Regular Public Transit Users

Yitong Chen[1], Wentao Dong[1], Chengcheng Yu[1,2], Quan Yuan[1,2] and Chao Yang[1,2,*]

1. Key Laboratory of Road and Traffic Engineering, Ministry of Education at Tongji University, 4800 Cao'an Road, Shanghai, 201804, China
2. Urban Mobility Institute, Tongji University,1239 Siping Road, Shanghai, 200092, China

**Abstract:** Data-driven research is becoming a new paradigm in transportation, but the natural lack of individual socio-economic attributes in transportation data makes research such as activity purpose inference and mobility pattern identification lack convincingness and verifiability. In this paper, a two-stage trip purpose and socio-economic attributes inference model is proposed based on travel resident survey and smart card data. In the first stage, the trip purpose of each trip is inferred by a combination of rule-based and XGBoost models. In the second stage, based on the trip purpose, a machine-learning model is built to inference the socio-economic attributes of individuals. A teacher-student model based on self-training is then applied on the models above to transfer them to smart card data. The impact of independent variables of socio-economic attributes inference model is also investigated. The results show that models for inferring trip purposes and socio-economic attributes have overall accuracies of 92.7% and 76.3%, respectively. Travel time, arrival time, departure time and purpose of the first two trips are most important factors on age and job status, while the land price of jobs-housing are significant to the deduce of individuals' incomes.

## 1 Introduction

Urban transit systems have been widely acknowledged to be essential to transportation sustainability in the way of alleviating the pressure of urban traffic congestion and environmental pollution. However, the current mismatch between the allocation of urban public transportation capacity and passenger demand has resulted in a waste of public resources, making it urgent to accurately grasp the characteristics of passenger demand (Yang *et al.*, 2023; Yu *et al.*, 2024a). In order to have a deep insight into people's travel mode choices and the complexity of travel purposes, a variety of data such as GPS data, smart card data, and social media data have been used to establish more precise disaggregated models, which have great value in understanding human mobility and travel behavior at both individual and aggregate levels (Barbosa *et al.*, 2018).

Although a mass amount of data from different source has been generated and collected from different sources, most of them lack information related to the trip purpose or subjects' socio-economic attributes, making these big data difficult to apply in travel behavior research. Researchers have found that socio-economic attributes have proven to play important roles in human travel behavior (Lenormand *et al.*, 2015). Attributes such as age (Shin, 2021; Wong *et al.*, 2018), gender (Simićević *et al.*, 2016), income (Di Ciommo *et al.*, 2014), occupational status (Silm *et al.*, 2024), and educational level (He *et al.*, 2020), have great influences on people's spatiotemporal characteristics and selection of travel modes. Meanwhile, activity-based models (ABM) were carried out to capture the nuances of individual activity-travel behavior and homogeneity and heterogeneity between individuals (Allahviranloo *et al.*, 2017; Rasouli & Timmermans, 2014). ABMs focus on individuals' mechanism of travel generation and lay an emphasis on trip purposes as a result (Yin *et al.*, 2021). On the other hand, manually collected datasets such as household survey data do have trip purpose and socio-economic attributes information, these data always have a relatively small quantity compared with those auto-collected data. If the advantages of the above two types of data are combined, we can get a large dataset with labeled trip purpose and socio-economic attributes, providing a better foundation to look into the relationship between travel demand and social and economic development.

In recent years, many approaches have been developed to deduce the socio-economic attributes of individuals in different kinds of data. Some researchers have proposed rule-based models (Alencar *et al.*, 2015), machine-learning models such Bayesian conditional probability model (Bi *et al.*, 2013), and deep-learning models (Kim *et al.*, 2022; Tkalčič & Ferwerda, 2018) to infer the socio-economic attributes of users. However, the above-mentioned research has three shortcomings. First, these models were trained and validated on the same dataset and failed to consider the method of transferring the model from a small dataset to a large dataset. Second, the travel purpose is not explicitly included in the model, making it hard to look into the relationships between socio-economic attributes and activities of individuals. Lastly, little research considered the interpretability of inference models, thus hindering the understanding of the relationship between travel patterns and socio-economic attributes.

Considering these shortcomings in the existing studies, this paper proposes a two-stage method for inferring both travel purposes and socio-economic attributes of regular transit users. The first stage is to infer the purpose of each trip based on resident travel survey data and POI data by using a combination of rule-based and machine-learning model, and a self-training teacher-student cycling strategy is built to transfer the model onto smart card data. Then, in the second stage, features including temporal and spatial travel characteristics, land price, population density, and trip purpose are used to build socio-economic attributes inference model. The self-training strategy is also applied in this part. Finally, the above models are compared with other models, and the effectiveness of the model has been validated. Some interpretability analysis has also been conducted. The contributions of this work lie in the following two aspects:

(1) Proposing a self-training model of travel purpose and socio-economic attributes based on public transportation, providing theoretical support for the annotation of large-scale unlabeled smart card data;

(2) Establishing a two-step joint inference framework of travel purpose and socio-economic attributes, analyzing the deep mechanism between socio-economic attributes and travel purposes and providing a new perspective for understanding the relationship between individuals' activities and their socio-economic attributes.

The rest of the paper is organized as follows. The second section presents a literature review on the inference of trip purpose and socio-economic attributes. The third section describes the datasets used in this paper, mainly the smart card dataset and the resident travel survey dataset, and the overall framework of inference of trip purpose and socio-economic attributes. The fourth section evaluates the effectiveness of the model. At last, the final section concludes the paper with an outlook.

# 2 Literature review

## 2.1 Trip purpose inference

At present, there have been many studies on the identification of travel purposes, and existing articles involve the identification of travel purposes from different data sources, including mobile GPS data (Cui *et al.*, 2018; Xiao *et al.*, 2016), taxi trajectory data (Gong *et al.*, 2015; Wang *et al.*, 2017), and bus swiping card data (Alsger *et al.*, 2018; Kusakabe & Asakura, 2014). Although the sources of data vary, most research methods use spatial geographic attributes near boarding and alighting, as well as the time of boarding and alighting, for identification.

In early research, there have been relatively few classifications of travel purposes in previous studies. Chakirov and Erath (2012) simply divided travel behavior into two categories: going to work and going home. Subsequently, Devillaine *et al.* (2012) further refined travel behavior into three categories: going to work, going to school, and going home. With the development of data mining technology, the accuracy of classification has improved, and the number of travel purposes has been further divided. Xiao *et al.* (2016) classified travel purposes into six categories based on mobile GPS data: picking up and dropping off people, social visits, shopping, dining, going to work (school), and returning home. Gong *et al.* (2015) proposed nine different travel purposes based on taxi trajectory data, including going home, working, transferring, eating, shopping, entertainment, going to school, accommodation, and seeking medical treatment.

In terms of modeling, Stopher *et al.* (2008) proposed a travel purpose inference method based on custom rules, which can successfully identify over 60% of travel records. Alsger *et al.* (2018) adopted an inference method based on conditional probability, and the accuracy of the discrimination method using spatiotemporal conditional probability can reach 78%. Chen *et al.* (2010) proposed a travel purpose recognition method based on the Logit model, proposing four types of travel purposes with discrimination accuracies of 67% and 78% for both home and non-home groups, respectively. Xiao *et al.* (2016) proposed a discrimination method based on artificial neural networks, which achieved a maximum accuracy of 97.22% on the test set. Sari Aslam *et al.* (2021) established ActiveNET model based on artificial neural networks and deep learning, using labeled data for prediction, with an overall accuracy rate of 94%.

## 2.2 Socio-Economic attributes inference

With the application of big data analysis in the field of transportation, many studies have begun to investigate individuals involved in transportation and classify them into different groups. Related research focused on two different directions: individual activity patterns (spatiotemporal characteristics) and socio-economic attributes. A large amount of research has been conducted on individual activity patterns (Cho *et al.*, 2023; Kim *et al.*, 2014; Ma *et al.*, 2013; Yu *et al.*, 2024b), while research on socio-economic attributes is quite few. Previous studies have shown that individuals' socio-economic attributes have strong impacts on their travel patterns (Hasan *et al.*, 2023; Pas, 1983). Therefore, it is necessary to establish accurate and effective models for identifying socio-economic attributes.

In previous studies, researchers have proposed various models for inferring socio-economic attributes based on data related to transportation. Solomon *et al.* (2018) used trajectory data to establish a Word2Vec model, extracted features from trajectory data, and predicted user attributes such as gender, age, and marital status. Results show that the proposed method achieves around 80% accuracy for various demographic prediction tasks. Kim *et al.* (2022) proposed a novel method to estimate the trip purpose and sociodemographic attributes of smart card data by using a conditional generative adversarial network (CGAN), reaching an overall accuracy of 68.55%. Wu *et al.* (2019) extracted spatiotemporal features from GPS trajectories and attached geographical context to trajectories to reflect people's activities. Supervised classification models are established and results show the inference accuracies of demographics range from 66% to 80%. Zhang *et al.* (2020) employed a multi-task convolutional neural network (CNN) on smart card dataset provided by Transport for London. It shows that the highest accuracies of car ownership, age group, income, and gender inference achieve 80%, 76%, 69%, and 64% respectively. Zhao *et al.* (2022) draws upon the Inverse Discrete Choice Modelling (IDCM) approach to enrich people's socio-demographic attributes. The accuracies vary from 50% to 90%. Zhang *et al.* (2024a) proposed a method first calculating the similarity of people's multidimensional daily activities and then applying dynamic time warping to measure the similarities of the multidimensional activity sequences. The results show that the proposed method outperforms existing methods in predicting four selected demographics: gender, age, education level, and work status, with an accuracy range between 91% and 94% for the national dataset and 88% to 91% for the GPS data. The summary of these existing studies for socio-economic inference is listed in Table 1.

Although so many different kinds of models have been proposed, few of them include the analysis of interpretability of models. Also, most of existing studies used GPS data, which achieved good inference accuracy due to its fine granularity, ranging from 66% to 94%. However, smart card data is relatively sparse in time and space, containing little information. Therefore, the accuracy of socio-economic attribute estimation based on smart card data is relatively low, basically not exceeding 80%.

In summary, the inference of trip purpose and socio-economic attributes is usually separated, and trip purpose is not considered when inferring socio-economic attributes. A joint inference model of travel purposes and socio-economic attributes needs to be proposed to reveal the relationship between them and improve the accuracy of the model.

**Table 1.** Summary of existing studies for socio-economic inference

| Reference | Data source | Method | Inferred attributes | Accuracy |
|---|---|---|---|---|
| Solomon *et al.* (2018) | GPS | Word2vec | Gender, age, marital status, academic faculty, if have children | 74%~87% |
| Wu *et al.* (2019) | GPS | XGBoost | Marital status, gender, residency status, education, age | 66%~82% |
| Kim *et al.* (2022) | Smart card data | CGAN | Income, car availability, driver's license, age, gender, home type | 68.55% |

| Zhang et al. (2020) | Smart card data | Multi-task CNN | Car ownership, age group, income, gender | 64%~80% |
| Zhao et al. (2022) | Household survey data | Inverse Discrete Choice Modelling | Income, car ownership, gender, driver's license, work status | 50%~90% |
| Zhang et al. (2024a) | GPS | Dynamic time warping | Gender, age, educational level, work status | 88%~94% |

## 3. Methodology

### 3.1 Data description

In this paper, the city of Shenzhen, China, is chosen as a case study for empirical research. As of 2018, Shenzhen's public transportation contains 981 regular bus lines, 5,331 pairs of bus stops, 21,000 km of regular bus operating line length, and an average daily passenger volume of 4.463 million. To establish an inference model for travel purposes and socio-economic attributes, resident travel survey data, smart card data, and Shenzhen's POI data are used in this research. The detailed descriptions of these three types of data are listed below.

**Resident Travel Survey Data. (RTSD)** The RTSD consists of one-day resident travel information obtained from a face-to-face survey conducted in the fourth quarter of 2020. Seeing that COVID-19 had already been brought under control in China at that time, its impact on people's travel was negligible. It is also noteworthy that all of the surveys were carried out on workdays. The survey questions were divided into three categories: personal socio-economic attributes (e.g., gender, age, income, job type, car ownership, education, home location, work location) and travel-related attributes (e.g., departure time, arrival time, departure location, arrival location, activity type, travel mode). The origin data contains trips of different travel modes but only trips by regular transit are used in order that we can focus on the travel of regular transit users. Finally, 20,802 trips of 5,996 users of regular public transit are selected in this dataset. The socio-economic information of these users and the purposes of the trips they took are summarized in Table 2. The raw data contains 11 types of travel purposes, which are categorized into 4 purposes that are most relevant to the socio-economic attributes. At the same time, some of the socio-economic attributes have been merged to reduce the complexity of the model.

**Smart Card Data. (SCD)** A sample of one-month regular transit SCD sample of Shenzhen from October 1st to October 31st, 2018 is utilized to validate the effectiveness of the model, including card swiping records of 12,979,574 users. Each trip contains the following fields: user ID, date, route, trip start and end time, boarding stop and alighting stop, trip start and end latitude and longitude. The data pre-processing eliminates some missing data and abnormal data, and transfer behaviors are also recognized and merged so that each record represents a trip instead of a ride. A survey revealed that more than 94% of the transfer activities took less than 60 min (Jang, 2010), which is taken as a threshold to represent a transfer activity between two rides in this paper.

**POI Data.** The POI data of Shenzhen in 2020 is obtained from Gaode Map API to add geospatial information to each trip.

**Population Data and Land Price Data.** The population data and land price data of Shenzhen in 2020 is obtained from WorldPop (https://hub.worldpop.org/) and Lianjia (https://sz.lianjia.com/) respectively.

Table 2. Descriptive statistics of socio-economic and trip purposes of RTSD

| Attributes | Category | Proportion (%) |
|---|---|---|
| **Socio-economic Attributes** | | |
| Gender | Male | 48.0 |
| | Female | 52.0 |
| Age | <20 | 17.3 |
| | 20-59 | 80.0 |
| | >=60 | 3.7 |
| Income | 0 | 11.8 |
| | 0-10 | 60.6 |
| | 10-15 | 19.8 |
| | >=15 | 7.8 |
| Job Status | With job | 69.8 |
| | Student | 15.6 |
| | Retired & No job | 14.6 |
| Car ownership | Yes | 32.7 |
| | No | 67.3 |
| **Trip Purpose** | | |
| Trip Purpose | Work | 41.4 |
| | Home | 48.5 |
| | Shopping & Entertainment (S&E) | 9.6 |
| | Medical | 0.5 |

### 3.2 Overall framework of the model

In this part, a two-stage method for inferring both travel purposes and socio-economic attributes of users of regular transit is proposed. The overall framework of the model is shown in Fig. 1. Due to the fact that the RTSD only includes one-day travel data for each resident, the residents with records of public transportation trips are more likely to be high-frequency users of public transportation. Moreover, the resident travel survey was carried out on weekdays, while the smart card data covers trips on both weekdays and weekends. The above two differences between RTSD and SCD make it difficult to establish a unified model on both datasets. Therefore, a semi-supervised self-learning-based method is proposed. In the first stage, both RTSD and SCD are divided into two categories based on whether the homes and workplaces of the users can be recognized and two initial models are established on RTSD. After that, by utilizing SCD for self-training, the trip purposes on SCD can be inferred. Then in the second stage, the predicted trip purposes are applied in the establishment of features of one-day trip chains and again a self-learning-based model is utilized to deduce socio-economic attributes on SCD. Finally, considering multiple days of travel chain data, the socio-economic attributes of users are determined using the maximum probability estimation method.

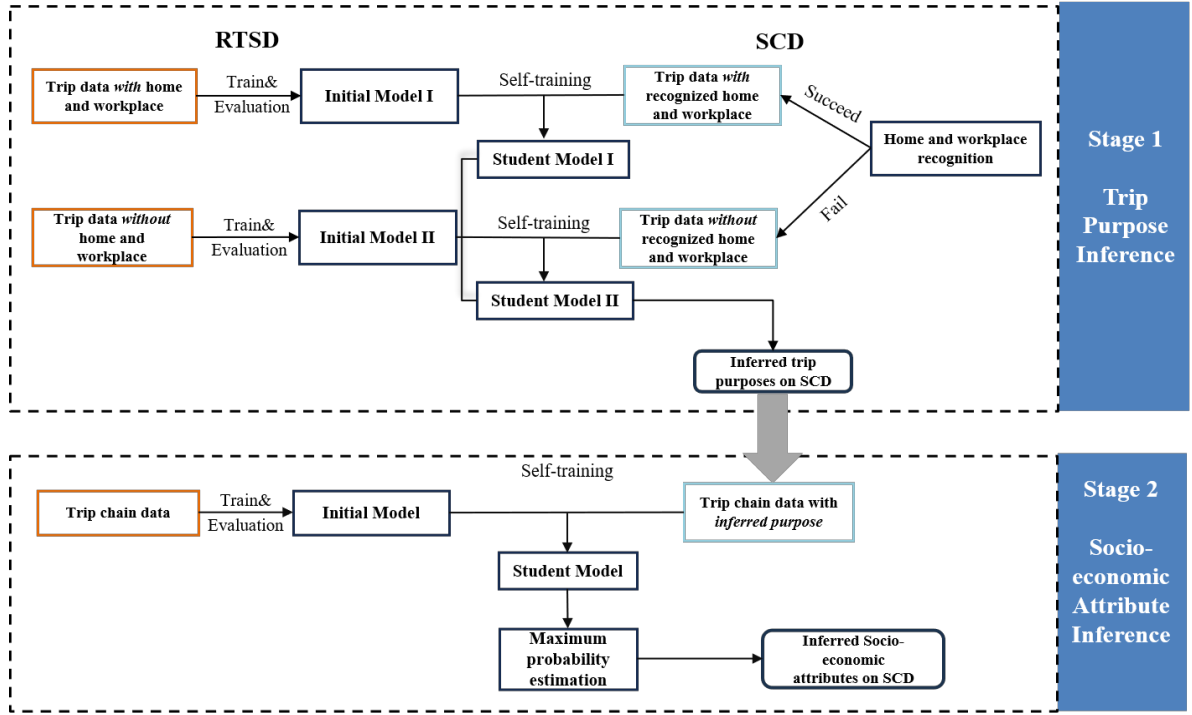

**Fig. 1** Overall Framework of Model

### 3.3 Initial Models for Inferring Trip Purposes

The Anchor Point Theory was proposed by Golledge R (1978), suggesting that the activity space of urban residents is framed by "anchor points" (frequently visited activity points such as homes and companies). The research pointed out that the frequency of trips between anchor points is much more than the frequency of other trips. As shown in the travel survey data of Shenzhen residents, 89.9% of trips are related to home or workplace. Therefore, process of inferencing trip purposes is shown in Fig. 2. Whether the location of the traveler's home and workplace can be identified is of crucial importance for deducing the purpose of travel. If a user's home and workplace can be identified, we first identify their work and home purposes based on rule-based model and then recognize the remaining purposes. Otherwise, all four purposes will be directly identified. In view of the lack of activity information between trips in SCD, the recognition of home and workplace can only rely on the regularity of boarding and alighting stops of SCD. The methods for identifying home and workplace are primarily based on the study by Ma *et al.* (2017) and have been improved upon since then. The rules of identifying home station $s_{Home}$ and workplace station $s_{Work}$ are shown by Equation (1) - (2).

$$s_{Home} = \arg\max_{s_{first} \in \{s_1, s_2, \ldots, s_n\}} f(s_{first}) \quad if \ f(s_{first}) > 0.6 \qquad (1)$$

$$s_{Work} = \arg\max_{s_{last} \in \{s_1, s_2, \ldots, s_n\}} f(s_{last}) \quad if \ f(s_{last}) > 0.6 \qquad (2)$$

where $s_{first}$ is the station where users first board buses in daily travels (departing before 3 p.m.), $s_{last}$ is the station where users last board buses in daily travels (departing after 3 p.m.).

Then, if the alighting stop is the workplace, the purpose of the trip will be identified as work. Similarly, if the alighting stop is home, the purpose of the trip will be identified as work. (corresponding to the rule-based model in Fig. 2)

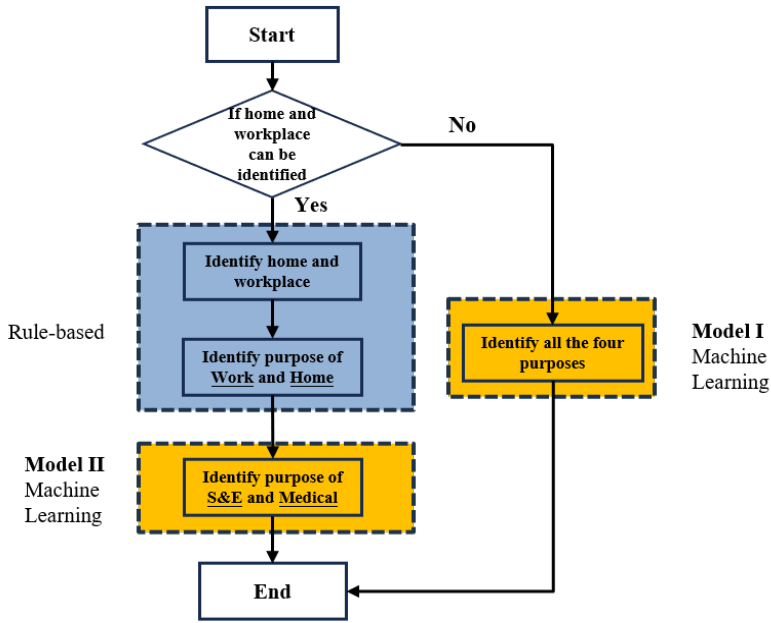

**Fig. 2** Diagram of Travel Purpose Identification Process

Then, when it comes to the trip purposes of the users whose home and workplace cannot be recognized (corresponding to model I in Fig. 2) or purpose of S&E and Medical for users who have identified home and workplace (corresponding to model II in Fig. 2), XGBoost model is applied. XGBoost is an improved gradient-boosting decision algorithm created by Chen and Guestrin (2016), which has high computational efficiency allowing for fast processing of large datasets, and its built-in regularization techniques help prevent overfitting and improve model generalization. The objective function of XGBoost is represented by Equation (3):

$$J(\theta) = L(\theta) + \Omega(\theta) \quad (3)$$

where L is the training loss function and $\Omega$ is the regularization function.

To solve the problem of imbalanced distribution of sample categories, a weighted cross entropy function is used as the loss function in the model. The detailed loss function and regularization function is shown by Equation (4) – (5):

$$L = -\frac{1}{N}\sum_{i=1}^{N}\frac{1}{n_{y_i}}\log(\hat{p}_{i,y_i}) \quad (4)$$

$$\Omega = \gamma T + \frac{1}{2}\lambda\sum_{j=1}^{T}\omega_j^2 \quad (5)$$

where $N$ represents the sample size, $n_{y_i}$ represents the number of samples belonging to category $y_i$, $\hat{p}_{i,y_i}$ represents the predicting probability of sample $i$ belonging to category $y_i$, $T$ represents the depth of trees and $\omega$ represents the vector of scores of leaves.

Finally, temporal features and land use features are included as the independent variables of the XGBoost model of inferencing trip purposes. The variables of boarding time and alighting time have been transformed into cosine values to capture the periodic features of the data (Reitboeck & Brody, 1969) and thus improve the model's generalization ability. The calculation method is shown by Equation (6):

$$COS_t = \cos\left(\frac{2\pi t}{T}\right) \qquad (6)$$

where t represents the seconds of time from 0:00 on that day, T represents the whole seconds of a day. The independent variables of the trip purpose model and their descriptions are illustrated in TABLE 3.

**Table 3.** Independent variables of trip purpose model

| Variable | Description |
|---|---|
| cos_boardtime | Cosine value of boarding time |
| cos_alighttime | Cosine value of alighting time |
| time_interval | Duration of the trip |
| board_catering | Proportion of catering POIs among all POIs at the boarding stop |
| board_education | Proportion of education POIs among all POIs at the boarding stop |
| board_leisure | Proportion of leisure POIs among all POIs at the boarding stop |
| board_shopping | Proportion of shopping POIs among all POIs at the boarding stop |
| board_hospital | Proportion of hospital POIs among all POIs at the boarding stop |
| board_company | Proportion of company POIs among all POIs at the boarding stop |
| board_residence | Proportion of residence POIs among all POIs at the boarding stop |
| alight_catering | Proportion of catering POIs among all POIs at the alighting stop |
| alight_education | Proportion of education POIs among all POIs at the alighting stop |
| alight_leisure | Proportion of leisure POIs among all POIs at the alighting stop |
| alight_shopping | Proportion of shopping POIs among all POIs at the alighting stop |
| alight_hospital | Proportion of hospital POIs among all POIs at the alighting stop |
| alight_company | Proportion of company POIs among all POIs at the alighting stop |
| alight_residence | Proportion of residence POIs among all POIs at the alighting stop |

### 3.4 Self-training model for Inferring Trip Purpose

Due to the previously mentioned significant differences between RTSD and SCD, a semi-supervised self-training model inspired by Gao *et al.* (2024) is utilized to detect the socio-economic attributes of users on SCD. Self-training is particularly useful in scenarios where there are limited labeled examples but an abundance of unlabeled data available for training (Amini *et al.*, 2022), and unlabeled data can help models learning by explicitly defining the structure of the data through a manifold (Niyogi, 2013). In this way, large-scale unlabeled smart card data can be utilized and realize generalization on smart card dataset.

The framework of self-training method is illustrated in Fig. 3. The model training process can be divided into two steps. The first step involves training an Initial Teacher Model (ITM) on RTSD using supervised learning methods. Then, the ITM is used to predict the activity purposes of trips in the SCD. When the confidence of the predictions exceeds set thresholds, the predicted types are applied to the unlabeled data as pseudo-labels, and the confidence threshold is denoted as $\tau$. Considering different learning status and learning difficulties of different classes, Curriculum Pseudo Labeling (CPL) (Zhang *et al.*, 2024b) is used to determine the threshold of labeling. The core of CPL is to flexibly adjust thresholds for different classes at each time step, thus ensuring informative and reliable unlabeled data and their pseudo labels. The calculation of confidence threshold T is illustrated by Equation (7) – (9):

$$\sigma_t(c) = \sum_{n=1}^{N} I(\max(p_{m,t}(y|u_n)) > \tau) \cdot I(\arg\max(p_{m,t}(y|u_n)) = c) \tag{7}$$

$$\beta_t(c) = \frac{\sigma_t(c)}{\max_c \sigma_t} \tag{8}$$

$$T_t(c) = \beta_t(c) \cdot \tau \tag{9}$$

where $p_{m,t}(y|u_n)$ the model's prediction for unlabeled data un at time step $t$, $N$ is the total number of unlabeled data, $\sigma_t(c)$ reflects the number of high-quality sample of class $c$ learned at time step $t$, and $\tau$ is the fixed threshold.

In our experiments, we observed that during the initial phase of training, the model tends to predict most unlabeled samples as belonging to a specific class, which is influenced by the parameter initialization. This behavior may cause a higher likelihood of confirmation bias, making the learning status estimates unreliable. Therefore, a warm-up process is introduced by rewriting the denominator in Equation (10) as:

$$\beta_t(c) = \frac{\sigma_t(c)}{\max\{\max_c \sigma_t(c), N - \sum_c \sigma_t(c)\}} \tag{10}$$

where the term $N - \sum_c \sigma_t(c)$ can be regarded as the number of unlabeled data that have not been used. This ensures the values of $\beta_t(c)$ increase gradually and prevent the model from categorizing all samples into the same type.

The second step involves the Teacher-Student Switching Model (TSSM), which is trained using a mix of labeled RTSD and pseudo-labeled SCD. Subsequently, the student model is used as new teacher model to generate predictions on the unlabeled SCD, which is used in the next training epoch. This alternating process is repeated until the model's performance on the new dataset reaches its best metrics. The ITM is trained on the RTSD and TSSM is developed on the SCD. The detailed process is shown in Algorithm 1.

**Algorithm 1.** Self-training teacher-student model training Process

1: **Input:** $D_{SCD} = \{(x_m, y_m), m \in (1, 2, ...M)\}$, $D_{RTSD} = \{u_n, n \in (1, 2, ..., N)\}$ (M labeled data and N unlabeled data.)

2: *ITM Training*

3: Create a supervised learning model as the initial teacher model $M_t$.

4: Initialize parameters $\theta_t$ of initial teacher model $M_t$.

5: Train $M_t$ by minimizing weighted cross-entropy loss (Eq. 3-5) on labeled dataset $D_{RTSD}$.

6: **while not converged do**

7:     $\theta_t \leftarrow \theta_t - \nabla\theta_t\, Loss(D_{RSTD}(\theta_t))$

8: **end while**

9: *TSSM Training*

10: **while** not reach the maximum iteration **do:**

11:     **for** c = 1 to C **do**

12:         Calculate $\sigma(c)$ using Eq. 7.

13:         **if** $\max \sigma(c) < \sum_{n=1}^{N} I(\hat{u}_n = -1)$ **then**

14:             Calculate $\beta(c)$ using Eq. 10.

15:         **else**

16:             Calculate $\beta(c)$ using Eq. 8.

17:         **end if**

18:         Calculate $T(c)$ using Eq. 9.

19:     **end for**

20:     Predict activity type on the unlabeled dataset $D_{SCD}$ by $M_t$. If the confidence level of predicted activity type $> T(c)$, the predicted activity type is adopted as 'pseudo-label'.

21:     Get $D_{mixed}$ by combining $D_{pseudo}$ with $D_{RTSD}$

22:     Create an equal model as a student model $M_s$.

23:     Initialize parameters $\theta_s$ of student model $M_s$.

24:     Train $M_s$ by minimizing weighted cross-entropy loss on $D_{mixed}$

25:     **while** not converged **do**

26:         $\theta_t \leftarrow \theta_t - \nabla\theta_t\, Loss(D_{mixed}(\theta_t))$

27:     **end while**

28:     **if** metrics perform better than the last iteration then

29:         close

30:     **else**

31:         Set the student model $M_s$ as the new teacher model $M_t$

32:         **goto** while

33:     **end if**

34: **end while**

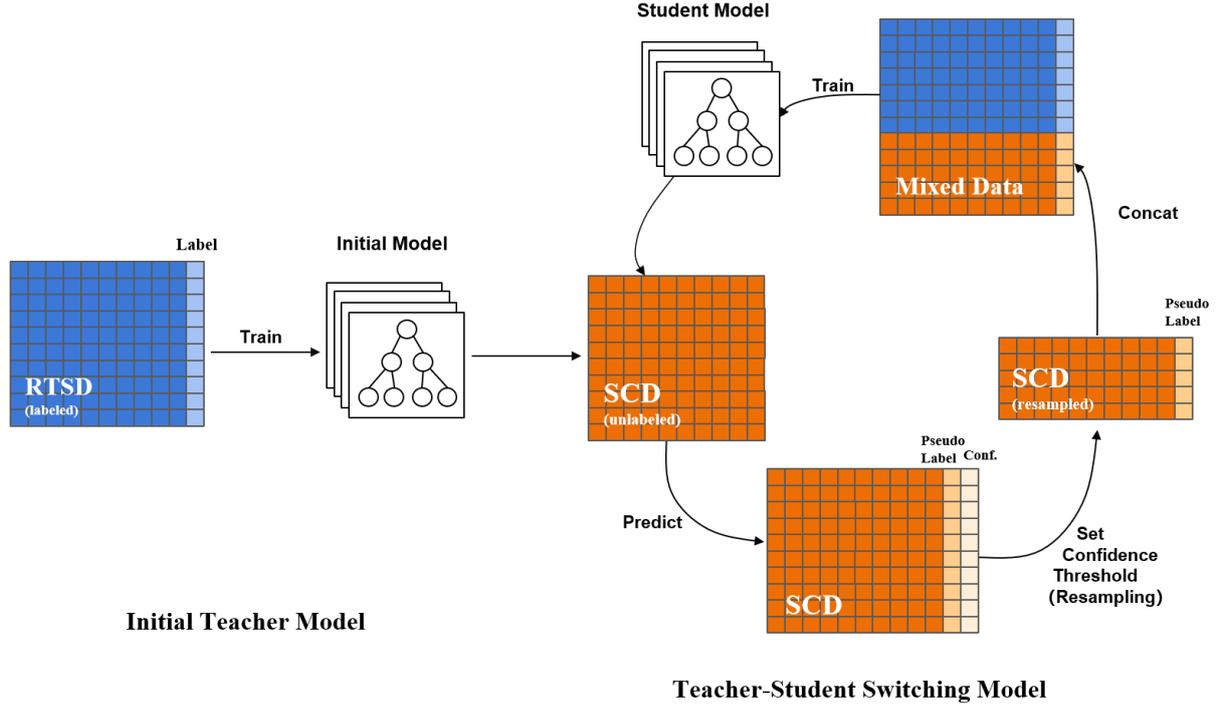

**Fig. 3** Framework of self-training method

### 3.5 Models for Inferring Socio-Economic Attributes

After finishing inferring the purpose of each trip, the model for inferring socio-economic attributes of regular transit users is proposed in this part. A lot of research involves the concept of social segregation, indicating that people who have the same socio-economic attributes tend to have similar temporal and spatial attributes of travel and live in similar communities, especially in big cities (Massey & Denton, 1988; Nilforoshan *et al.*, 2023). Due to the fact that resident survey data only contains one-day travel information of a single user, we can only obtain the relationship between daily travel and the user's socio-economic attributes. Therefore, a daily travel chain is established to infer the user's socio-economic attributes. As more than 99% of transit users make 5 or fewer trips in one day in RTSD, the spatial-temporal information about users' first 5 trips in one day are selected in the feature engineering (if a user traveled fewer than 5 times in a day, the variables of the unfinished trips will be set as -1). The variables about spatial information of user's home and workplace, such as population density and land price, are also included. The space area of Shenzhen city is divided into 1km*1km grids and numbered. The departure and arrival position of each trip is represented by the grid number and coded as cosine values, which is shown as Equation (11) and (12):

$$COS_x = \cos\left(\frac{2\pi x}{X}\right) \quad (11)$$

$$COS_y = \cos\left(\frac{2\pi y}{Y}\right) \quad (12)$$

where x, y represents the number of the grid on x-axis and y-axis respectively, X represents the total number of grids on x-axis and y-axis. The variables of socio-economic attributes model are shown in TABLE 4.

After the feature engineering, a socio-economic attribute inference method similar to travel purpose inference is built. All RTSD is firstly put into the XGBoost model for training, and then pseudo labels are generated on SCD based on self-training strategy (mentioned in the previous chapter). Considering confidence threshold control, the mixed data is used for the next round of model training to obtain socio-economic attribute inference model based one-day travel chains.

Although socio-economic attributes are inferred based on one-day trip chains, we wonder to get the real socio-economic attributes of regular transit users. Thus, a rule is proposed that the user's real socio-economic attributes are the most frequently appeared ones among all the attributes inferred during the long term, as is shown in Fig. 4. Assuming that the multi-day inferred travel purposes of user $i$ are $C_i = \{c_i^1, c_i^2, ..., c_i^n\}$, the real socio-economic attribute is defined by Equation (13):

$$c_i^{real} = \arg\max_{c \in C} f(c_i) \qquad (13)$$

where $f(c_i)$ represents the occurrence number of $c_i$ in set $C$. In this way, the real socio-economic attributes of users can be uniquely determined.

**Table 4.** Variables of socio-economic attributes model

| Category | Variable | Description |
|---|---|---|
| **Independent Variables** | | |
| **Spatial Information of Home (4)** | Home_X | Home's X cosine value of the grid |
| | Home_Y | Home's Y cosine value of the grid |
| | Home_Pop | Population density of home location grid |
| | Home_LP | Land Price of home location grid |
| **Spatial Information of Workplace (4)** | Work_X | Workplace's X cosine value of the grid |
| | Work_Y | Workplace's Y cosine value of the grid |
| | Work_Pop | Population density of workplace location grid |
| | Work_LP | Land Price of workplace location grid |
| **Information of Trips (11*5)** | Dep_Time1 | Departure time of the 1st trip of the day (cosine value) |
| | Arr_Time1 | Arrival time of the 1st trip of the day (cosine value) |
| | Travel_Time1 | Travel time of the 1st trip of the day |
| | Dep_X1 | X cosine value of the grid of 1st trip's departure location |
| | Dep_Y1 | Y cosine value of the grid of 1st trip's departure location |
| | Arr_X1 | X cosine value of the grid of 1st trip's arrival location |
| | Arr_Y1 | Y cosine value of the grid of 1st trip's arrival location |
| | Purpose1_1 | If the trip purpose of 1st trip is Work |
| | Purpose1_2 | If the trip purpose of 1st trip is Home |
| | Purpose1_3 | If the trip purpose of 1st trip is S&E |
| | Purpose1_4 | If the trip purpose of 1st trip is Medical |
| | …… | …… |
| | Dep_Time5 | Departure time of the 5th trip of the day (cosine value) |
| | Arr_Time5 | Arrival time of the 5th trip of the day (cosine value) |

| | Travel_Time5 | Travel time of the 5th trip of the day |
| | Dep_X5 | X cosine value of the grid of 5th trip's departure location |
| | Dep_Y5 | Y cosine value of the grid of 5th trip's departure location |
| | Arr_X5 | X cosine value of the grid of 5th trip's arrival location |
| | Arr_Y5 | Y cosine value of the grid of 5th trip's arrival location |
| | Purpose5_1 | If the trip purpose of 5th trip is Work |
| | Purpose5_2 | If the trip purpose of 5th trip is Home |
| | Purpose5_3 | If the trip purpose of 5th trip is S&E |
| | Purpose5_4 | If the trip purpose of 5th trip is Medical |
| **Dependent Variables** | | |
| **Socio-Economic Attributes (3)** | Age | Age of the individual |
| | Job Status | Job status of the individual |
| | Income | Income of the individual |

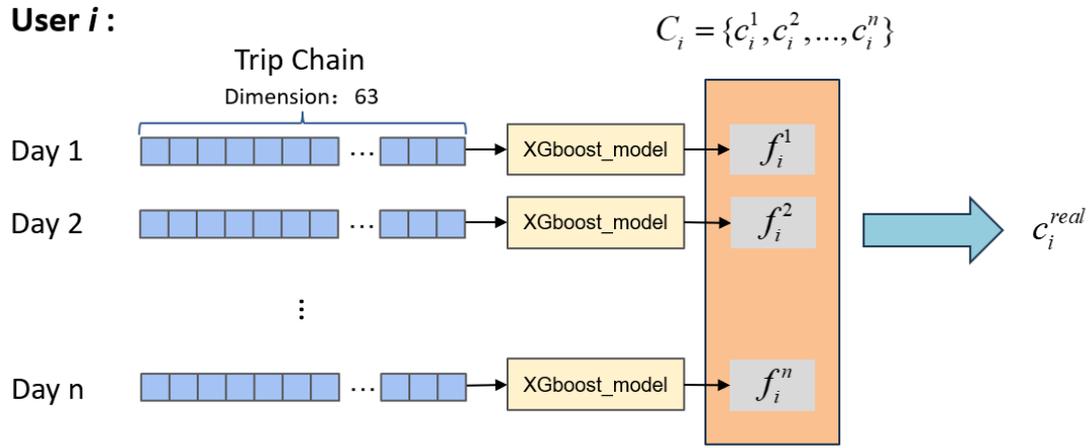

**Fig. 4** Socio-economic attribute inference based on multi-day travel data

### 3.6 Evaluation metrics

During the training process, both RTSD and SCD were divided into 80% training set and 20% testing set. the selection of hyperparameters of XGBoost is determined by the grid search method. For initial training and teacher-student switching process, we have designed different evaluation metrics to evaluate the performance of the models. In the process of initial training, accuracy, precision, and recall are considered. The metrics are calculated by Equation (14) – (16):

$$Accuracy = \frac{tp + tn}{tp + fp + tn + fn} \quad (14)$$

$$Precision = \frac{tp}{tp + fp} \quad (15)$$

$$Recall = \frac{tp}{tp + fn} \quad (16)$$

where $tp$ is the number of true positive, $fp$ is the number of false positive, $fn$ is the number of false negative, $tn$ is the number of true negative. The overall metrics is calculated by Equation (17):

$$Weighted\ metric = \sum_{i=1}^{N} \left( \frac{metric_i * n_i}{\sum_{i=1}^{N} n_i} \right) \tag{17}$$

where $N$ represents the number of classes, $n_i$ represents the sample numbers of class $i$, $metric_i$ represents the accuracy, precision, or recall of class $i$.

In view of the lack of socio-economic labels on SCD, Silhouette Coefficient and Davies-Bouldin Index are utilized as the metric to evaluate the performance of the model. For the discrimination of travel purposes, the same travel purpose has similar spatiotemporal characteristics. For the discrimination of socio-economic attributes, people with the same socio-economic attributes share similar characteristics of travel, residence, and work based on the social segregation theory. Thus, the above indicators can reasonably evaluate the effectiveness of the model. The Silhouette Coefficient is used to evaluate the quality of clustering by highlighting the separation and cohesion between samples, while the Davies-Bouldin Index focuses on internal divergence and inter-cluster separation at a cluster level. The calculations of the two metrics are illustrated by Equation (18) – (19):

$$S(i) = \frac{1}{n} \sum_{i=1}^{n} \frac{b(i) - a(i)}{\max\{a(i), b(i)\}} \tag{18}$$

$$DB = \frac{1}{k} \sum_{i=1}^{k} \max_{j \neq i} \left( \frac{\sigma(i) + \sigma(j)}{d(i,j)} \right) \tag{19}$$

where $a$ represents the average distance from other samples in the same cluster, $b$ represents the average distance from the cluster where the nearest sample is located, $n$ represents the number of samples, $\sigma(i)$ represents the average distance from samples within cluster $i$ to the centroid of the cluster, $d(i,j)$ represents the distance between the centroids of cluster $i$ and cluster $j$, $k$ represents the number of samples. The range of values for the Silhouette Coefficient is between -1 and 1, with the value closer to 1, the better the model. And if the Davies-Bouldin Index is smaller, the classification effect of model is better.

## 4. Results

### 4.1 Results for Inferring Trip purposes

The confusion matrices of the two machine-learning models are shown in Fig. 5. Fig. 5(a) and 5(b) show the confusion matrices of Model I and Model II respectively. The average accuracies of Model I and Model II are 86.7% and 78.3%. In Model I, the purpose of Work and

Home are well recognized, achieving an accuracy of around 90%, while the accuracies of the purpose of S&E and Medical are relatively low, at only 57.0% and 29.4%. The average accuracy of Model I is 78.3%. In Model II, the purpose of Medical is also not very well predicted, with an accuracy of only 47.3%. It seems that both Model I and Model II tend to classify purpose of Medical as purpose of S&E. Thus, the low accuracies for S&E and Medical may be due to the fact that the two purposes are usually dispersed in time and space with no clear boundaries, which is not conducive to the recognition of the model. Although the accuracies of the above two categories are not very high, they are minor travel purposes and have limited impact on inferring socio-economic attributes. Considering that rule-based models are generally accurate for the purpose of returning home and going to work, the accuracy of identifying the purpose of home and work is over 95% and the overall accuracy of our proposed trip purpose inference model in this paper is 92.7%.

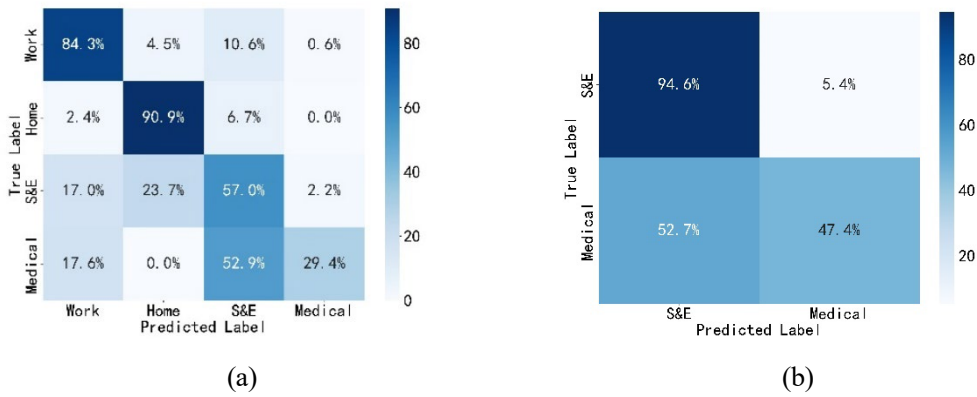

(a)          (b)

**Fig. 5** Confusion matrices of the two models on testing set

Further attention to the metrics between our proposed model and other machine learning models. The performance of the three models is shown in TABLE 4. As can be seen from the table, apart from our proposed model, XGBoost outperforms the other machine-learning models, which provides us with the reason for choosing it as the base model in Model I and Model II. Moreover, our proposed model outperforms other models in all evaluation metrics with accuracy increasing by 14.4%. This result demonstrates the effectiveness of identifying users' homes and workplaces first. By pre-identifying the locations of users' homes and workplaces, our proposed model can effectively distinguish between primary activities such as returning home and going to work, as well as other secondary activities.

**Table 4.** Performance comparison between proposed model and other models (Unit: %)

| Model | Accuracy | Precision | Recall |
|---|---|---|---|
| Proposed model | **92.7** | **91.2** | **92.7** |
| XGBoost | 78.3 | 77.4 | 78.3 |
| SVM | 68.4 | 65.6 | 68.4 |
| KNN | 71.2 | 69.7 | 71.2 |
| RF | 72.8 | 67.2 | 72.8 |
| LightGBM | 74.9 | 73.3 | 74.9 |

Note: The "proposed model" in the table refers to the metrics of the joint rule-based model and machine learning model.

In SCD, the home and workplace of users are firstly recognized. The successful rate of recognizing both home and workplace is 14.5%. Both model I and model II are then trained through self-training strategy. The results are shown as TABLE 5. The results show that the Silhouette Coefficient and the Davies-Boulding Index of TSSMs are all better than the ITMs, indicating self-training strategy improves the ITMs' performance of the trip purpose inference tasks on SCD. Then we predict the purpose of each trip and see whether the distributions of purposes on those datasets are similar. Given the RTSD are all on weekdays, SCD is divided into weekday dataset and weekend dataset to look into the differences. The results are shown in Fig. 6. We can see that the distribution of trip purposes on the SCD-weekday is more similar to that of RTSD than on the SCD-weekend. The proportion of S&E purposes on weekends is higher than proportions on the other two datasets, while the proportion of Work purpose is lower. This is because people tend to have more entertainment on non-working days. The proportion of S&E in SCD-weekday is also much higher compared to RTSD. This may be due to the fact that the respondents in RTSD are mostly high-frequency users, while SCD-weekday covers a large number of low-frequency users, whose travel purposes are mostly non-commuting. As a whole, although the distribution of travel purposes has subtle differences between the two sets of data, the overall distribution remained consistent.

**Table 5.** Comparison between trip purpose ITM and TSSM on SCD

|  | Model | Silhouette_Coef | DB_Index |
|---|---|---|---|
| Model I | ITM | 0.156 | 2.938 |
|  | TSSM | **0.173** | **1.672** |
| Model II | ITM | 0.049 | 18.186 |
|  | TSSM | **0.054** | **7.839** |

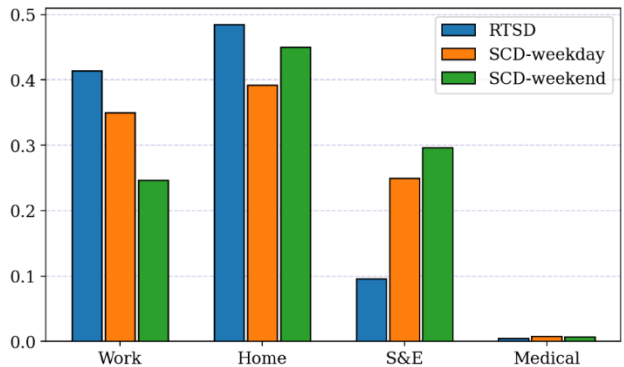

**Fig. 6** Trip purpose distribution on different datasets

### 4.2 Results for Inferring Socio-Economic Attributes

The confusion matrices of initial models of socio-economic attributes inference are shown in Fig. 7. Fig. 7(a), (b) and (c) show the confusion matrices of age, job status and income models respectively. Among all the three models, the job status model performs the best, with all three categories' accuracies higher than 70%. For age model, people aged under 20 and 20-59 are well predicted, with accuracies at 71.6% and 89.3% respectively. However, people aged above 60 are not clearly identified, which may be due to small sample size and the behavioral differences within the elderly categories. For income model, people who don't make money and those who earned 0-100 thousand yuan per month are fairly well recognized, with accuracies at the levels of 62.1% and 66.5%, respectively. The other two categories' accuracies are both lower than 50% and the average accuracy of income is merely 59.5%, indicating that the relationship between income and behavior of taking public transit is relatively weak.

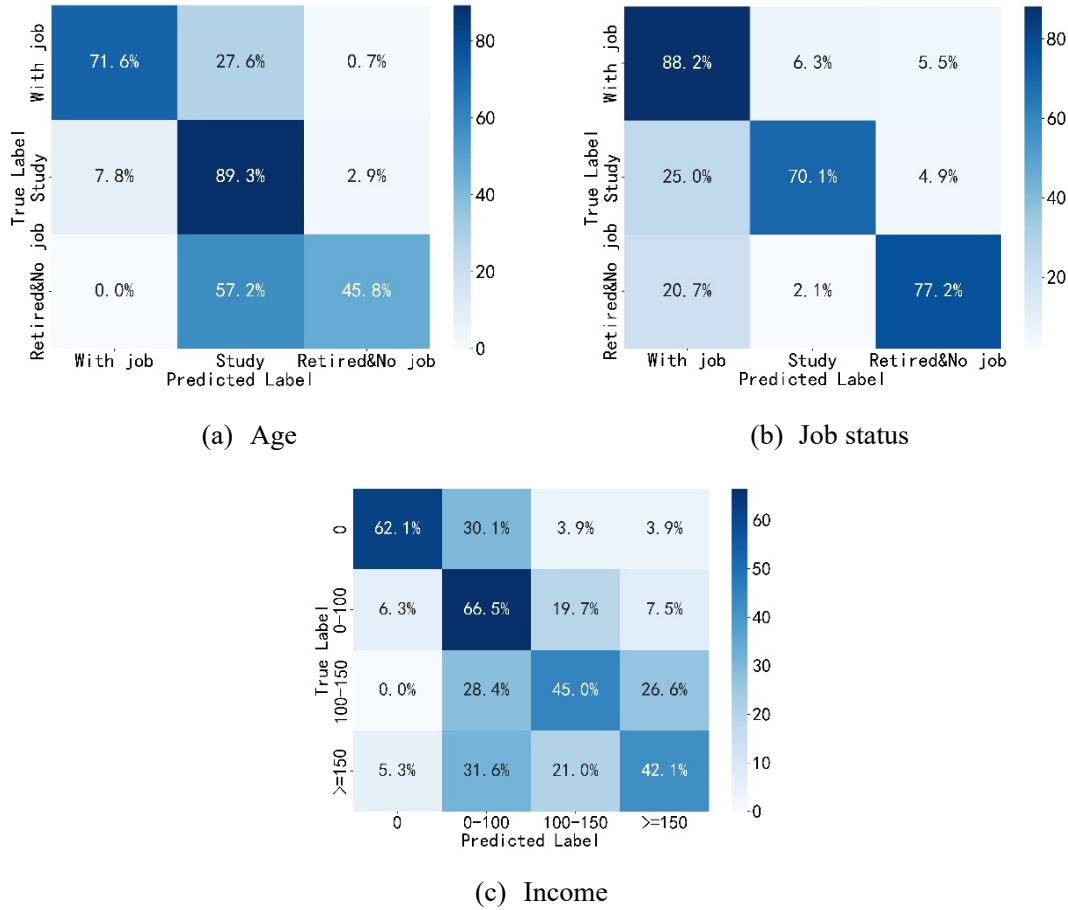

(a) Age

(b) Job status

(c) Income

**Fig. 7** Confusion matrices of socio-economic inference models on testing set

The performances of the initial teacher XGBoost models are compared with other models, including Support Vector Machine (SVM), k-Nearest Neighbor (KNN), Random Forest (RF), and LightGBM, across various attributes. Table 6 presents the results, which indicate that XGBoost consistently outperforms the other models in most cases, achieving overall accuracies of 85.5% and 83.8% for predicting age and job status, respectively, though its performance in inferring income was slightly lower at 59.5%. Despite lower accuracy and recall for income prediction compared to RF and LightGBM, XGBoost exhibited notably higher precision. The model achieved an average accuracy of 76.3% across all three attributes, demonstrating strong performance in inferring socio-economic characteristics from public transit trip data alone.

**Table 6.** Performance comparison of socio-economic inference models (Unit: %)

| Attributes | Model | Accuracy | Precision | Recall |
|---|---|---|---|---|
| | SVM | 81.4 | 76.4 | 81.4 |
| | KNN | 84.3 | 81.8 | 84.3 |
| Age | XGBoost | **85.5** | **88.2** | **85.5** |
| | RF | 85.3 | 85.0 | 85.3 |
| | LightGBM | 85.2 | 84.3 | 85.2 |
| Job Status | SVM | 79.1 | 76.2 | 79.1 |

|  | KNN | 81.7 | 80.5 | 81.7 |
|  | XGBoost | **83.8** | **85.3** | **83.8** |
|  | RF | 83.5 | 84.1 | 83.5 |
|  | LightGBM | 83.4 | 83.1 | 83.4 |
|  | SVM | 59.5 | 51.1 | 59.5 |
|  | KNN | 56.5 | 50.7 | 56.5 |
| Income | XGBoost | 59.7 | **71.8** | 59.7 |
|  | RF | **63.1** | 56.5 | **63.1** |
|  | LightGBM | 60.1 | 55.5 | 60.1 |

After verifying the effectiveness of the initial model, the performance of TSSM is also investigated. The model is tested on RTSD and the results are shown as TABLE 7. For all three types of socio-economic attribute inference models, their self-trained models achieved higher Silhouette Coefficients and lower Davies-Boulding Indexes The results indicate that the self-training strategy provides a good transferability, making the initial model's performance improve on SCD.

The comparison of socio-economic attribute distributions on two datasets is shown in Fig. 7. Although the distribution of socio-economic attributes is roughly the same, there are still some subtle differences. The age distribution is generally consistent between the two datasets, with slightly fewer individuals under 20 years old in SCD and a somewhat higher proportion of those over 60. The number of retired individuals is similar across both datasets, but there is a lower proportion of employed people in SCD, while the students are overestimated in SCD. With regard to income, there is a significantly larger proportion of individuals with incomes over 150,000, while those earning less than 150,000 are underestimated. This could suggest that SCD includes more low-frequency public transit users, who tend to have relatively higher incomes and use public transportation only occasionally.

**Fig 7.** Comparison between Socio-economic ITM and TSSM on SCD

| Attributes | Model | Silhouette_Coef | DB_Index |
| --- | --- | --- | --- |
| Age | ITM | -0.079 | 11.514 |
|  | TSSM | **-0.001** | **7.644** |
| Job status | ITM | -0.039 | 9.972 |
|  | TSSM | **-0.012** | **8.124** |
| Income | ITM | -0.020 | 20.074 |
|  | TSSM | **-0.007** | **5.922** |

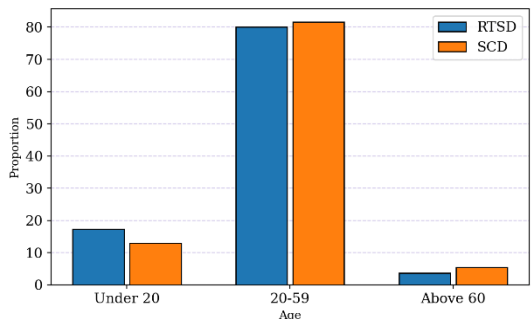

(a) Age

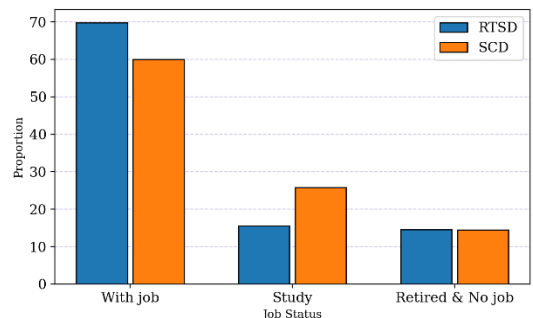

(b) Job status

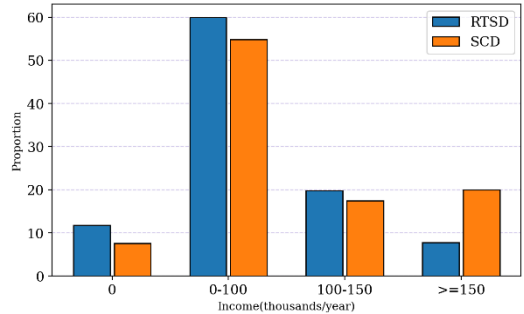

(c) Income

**Fig. 8** Socio-economic attributes distribution on different datasets

## 5. Discussion

### 5.1 Factors Influencing the Inference of Socio-Economic Attributes

In order to see the different influences of spatiotemporal, job-housing and trip purpose features on deducing models of socio-economic attributes, the average impacts of independent variables are looked into by SHapley Additive exPlanations (SHAP) (Lundberg & Lee, 2017). The larger the absolute SHAP value of an independent variable is, the greater its impact is on the model's dependent variable. The SHAP values of the three models are in Fig. 8, the spatiotemporal, job-housing and trip purpose features marked as different colors and their actual mean absolute SHAP value are noted under the variable name.

For age inference model, as shown in Fig. 9(a), the arrival and departure time of the first two trips in one day plays the most important roles in the recognition of people aged under 20 and between 20 to 59. Besides, if the first trip purpose being Home or S&E is also important for people aged under 20. This is interesting because the purpose of the first trip being Home indicates that the individual makes another trip without using regular transit before the first ride. We find that 9.2% of users in SCD have a first trip for Home purpose, and most of the trips are in the afternoon. This is probably because some parents drive their children to school in the morning and the children choose to take a bus home themselves in the afternoon. And for some college student aged under 20, they may have more freedom in travelling and have their first trip of a S&E purpose. On the contrary, the jobs-housing variables values the most in the recognition of people aged above 60. A missing work address will result in the user likely to be considered over 60 years old, and the residential addresses of the elderly also show a clustering phenomenon in space.

For job status model, as shown in Fig. 9(b), SHAP values illustrate similar results to the age model. The recognition of people who are with job and at school are mostly influenced by the temporal features of the first two trips, while the people who are retired depends on the employment situation and if the first purpose being S&E. Compared with the age model, the job status model has a more obvious relationship with trip purposes. Finally, for models of income, shown in Fig. 9(c), the land price of jobs-housing makes valuable contributions to the deduce of individuals' incomes compared with other two models. For those who make money every month, the land price of their home location may indicate their income status. And for those whose monthly incomes are over 10 thousand yuan, the Y coordinate of grid of home is also important, showing that the relative position of home in the city is related to a person's income. The analyses above are of great significance for understanding the underlying mechanisms of models and providing an insight into relationships between people's travel and socio-economic attributes.

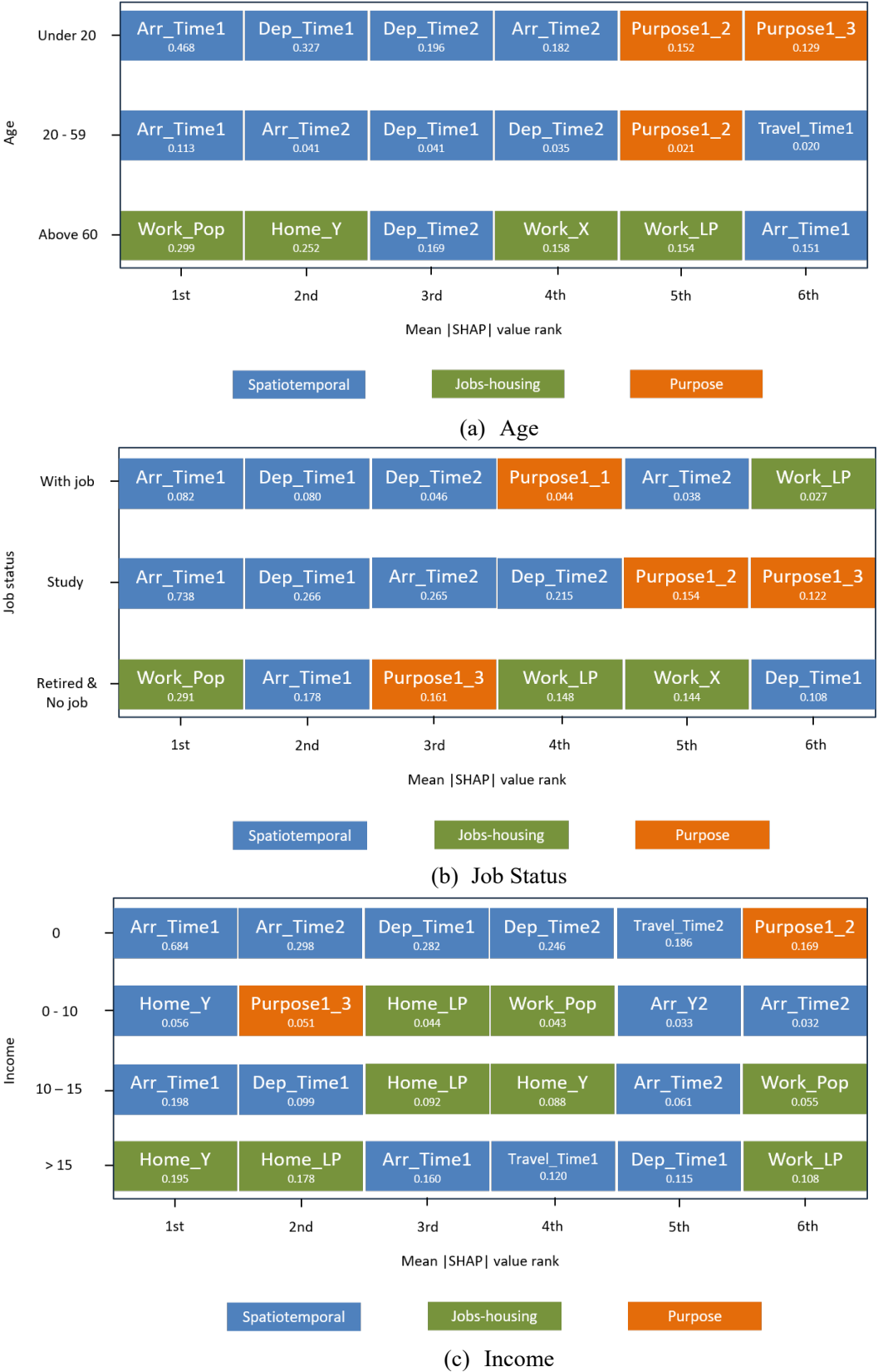

**Fig. 9** SHAP value of socio-economic inference model

# 6. Conclusion

The primary objective of this paper is a two-stage method for inferring both trip purposes and socio-economic attributes. A rule-based and machine-learning combined model is established to infer trip purposes and a machine-learning model is built to infer socio-economic attributes such as age, job status, and income. A self-training teacher-student switching model is also established to transfer the model from resident travel survey data to smart card data. The interpretability of the socio-economic model is also looked into. From the results, some major conclusions can be drawn. First, models for inferring trip purposes have an overall accuracy of 92.7%. The rule-based model can improve recognition performance for purposes of Work and Home, achieving an accuracy of over 95%. The proportion of trips of shopping and entertainment purposes in smart card data are higher than that in resident travel survey data. Second, models for inferring trip purposes have an overall accuracy of 76.3%. The accuracies of inferring age, job status, and income are 85.5%, 83.8%, and 59.7% respectively. The self-training framework can bring the model good transferability. Finally, travel time, arrival time, and departure time of the first two trips in a day of individual play the most important roles in recognizing all three socio-economic models. If the first trip of purpose being Home is also significant to identify students or individuals who are under 20, and the land price of jobs-housing makes valuable contributions to the deduce of individuals' incomes.

Despite the contributions of this study, several limitations must be acknowledged. First, due to the lack of smart card data of subway, only the regular bus trips are considered, which may neglect the transfer behavior between subway and bus and affect the accuracy of describing travel behaviors. Second, this method was only experimented on data from Shenzhen, which may limit the generalizability of the model to other urban environments. Experiments can be conducted on datasets from more cities in the future to demonstrate the generality of the model.

## Author Statement

*Yitong CHEN:* Research Design, Writing original draft, Data analysis, Visualization and Manuscript Draft Preparation. *Wentao DONG*: Data analysis and experiments, Visualization. *Chengcheng YU*: Editing, and Manuscript Draft Preparation. *Quan YUAN:* Editing, and Manuscript Draft Preparation. *Chao YANG*: Data Access provision, Conceptualization, Research Design, and Editing.

## Declaration of Competing Interest

The authors report no declarations of competing interest.

## Acknowledgment

The authors appreciate the funding support from NSFC grants (Grant No. 52172305).